\documentclass{desyproc}

\begin{document}
\title{Constraining ALPs with linear and circular\\Polarisation Measurements of Quasar Light}

\author{{\slshape Alexandre Payez}\\[1ex]
IFPA group, AGO Dept., University of Li\`ege, Li\`ege, Belgium\\
Deutsches Elektronen-Synchrotron (DESY), Hamburg, Germany
}

\contribID{familyname\_firstname}

\desyproc{DESY-PROC-2013-XX}
\acronym{Patras 2013} 
\doi  

\maketitle

\begin{abstract}
	We discuss the constraints derived on the mixing of photons with light pseudoscalars using the distributions of good-quality linear and circular polarisation measurements of light from the least polarised classes of quasars.
	We also provide the dependence of our limit on the average electron density in the local supercluster for nearly massless particles.
\end{abstract}


\section{Introduction}

	In theoretical extensions of the Standard Model, all the new particles are not necessarily on the heavy side; the prediction of very light (sub-eV) scalar or pseudoscalar particles that are very weakly interacting is actually quite generic. Looking for signatures of such ``axion-like particles'' (or ALPs) therefore represents another exciting possibility to probe the kind of physics there can be at very high scales, associated with these new degrees of freedom, which is complementary to collider searches.
	It is thus not surprising that a lot of ongoing effort is made to scan the parameter space of these hypothetical particles, actively searched for mostly via their electromagnetic coupling.
	Current experimental developments include for instance light-shining-through-a-wall experiments such as ALPS, which is being upgraded~\cite{Bahre:2013ywa}, or projects of next-generation helioscopes such as the International Axion Observatory~\cite{Vogel:2013bta}.

	Due to their mixing with photons, these spinless particles are moreover not only of interest in particle physics but also in astrophysics, as their existence would change the properties and the propagation of light emitted from distant sources. A sizeable signal could be expected as the distances involved are huge, even if the ALPs are only coupled very weakly. In this context, various phenomena have in fact already been interpreted as possible astrophysical hints for the existence of ALPs, especially in the low-mass region of the parameter space: namely ALPs with masses $m\lesssim 10^{-9}$~eV and couplings to photons $g\sim10^{-12}$--$10^{-11}$~GeV$^{-1}$. For a recent summary of the motivations for these particles from theory and of their implications, see Refs.~\cite{Payez:2013lwa,Baker:2013zta,Ringwald:2012hr} and references therein.

	In this work~\cite{Payez:2012vf}, we consider light coming from quasars. The fact that these high-luminosity active galactic nuclei (AGN) are among the brightest and furthest steady light sources known in the Universe makes them particularly appealing to look for signals of ALPs and to constrain their parameter space.
	We focus on the window of astrophysical interest and show that constraints can be derived using both linear and circular polarisation data.

\section{What polarisation can tell us}

	Polarimetry is an extremely valuable tool to search for axion-like particles as the interaction of these particles with photons, in external magnetic fields for instance, would change the polarisation of light~\cite{Maiani:1986md,Raffelt:1987im,Harari:1992ea}.
	Suitable interaction Lagrangians read
	\begin{equation}
		\begin{split}
				\mathcal{L}_{\phi\gamma\gamma} &= {\frac{1}{4}\ g \phi F_{\mu\nu}\widetilde{F}^{\mu\nu}}=-g\phi(\vec{E}\cdot\vec{B}) \textrm{ for pseudoscalar } \phi,\\
				\textrm{and }\mathcal{L}_{\phi\gamma\gamma} &= {\frac{1}{4}\ g \phi F_{\mu\nu}{F}^{\mu\nu}}=\frac{1}{2}\ g\phi({\vec{B}}^2 - {\vec{E}}^2) \textrm{ for scalar } \phi,\label{eq:AP_lagrangians}
		\end{split}
	\end{equation}
	and lead to similar phenomenological consequences; henceforth we focus on pseudoscalars.
	In a nutshell, one expects from the mixing that even unpolarised light will develop a non-vanishing degree of linear polarisation and that, in general, linear and circular polarisation will convert at least partially into one another as a consequence of phase-shifts induced by the mixing.\footnote{There are different regimes, with distinctive properties; more details can be found for instance in Ref.~\cite{Payez:2013lwa}.}
	As is well-known and readily seen from Eq.~\eqref{eq:AP_lagrangians}, this is because, in an external magnetic field, such spinless particles only couple to one direction of polarisation~\cite{Sikivie:1983ip}.

	The mixing can actually be very efficient at modifying polarisation and can therefore be constrained by precise measurements. The limits derived from the absence of rotation of the linear polarisation of UV light from AGN~\cite{HornsAlighieri} are a recent example of this.

	In this work, we follow a different idea, which is to consider the spectroscopically defined quasar classes known to have the smallest intrinsic polarisations in visible light, and to compare the predictions of the mixing with observations.
	As already discussed by Harari and Sikivie in Ref.~\cite{Harari:1992ea}, light from distant sources should be characterised by at least some amount of polarisation if axion-photon mixing happens along the way.
	We do this for both linear ($p_{\mathrm{lin}}$) and circular ($p_{\mathrm{circ}}$) polarisation.

	Quasar polarisation measurements at optical wavelengths can be found in the literature with uncertainties below 0.1\%. Light from quasars is known to be intrinsically linearly polarised, as differences can be seen in different spectroscopic types, and the polarisation is at the 1\%-level for the least polarised ones. If there are many catalogs of linear polarisation measurements, the same cannot be said about circular polarisation: it has rarely been studied, despite being measured in the same way as the linear one by simply adding a quarter-wave retarder plate.
	Following the predictions of the mixing, new dedicated observations of quasar circular polarisation in visible light have therefore been carried out~\cite{Hutsemekers:2010fw}. For observational reasons, most of the objects are located towards the North Galactic Pole direction, which points to the center of the local supercluster; see \textit{e.g.}~\cite{Courtois:2013yfa}.
	No evidence for non-vanishing circular polarisation could be found at the 3$\sigma$-level however~\cite{Hutsemekers:2010fw}.\footnote{That is, except for some highly linearly polarised blazars which could be intrinsically circularly polarised: as for radio waves, their optical emission would mostly be due to beamed synchrotron radiation from ultra-relativistic electrons rather than to the usual thermal emission from the accretion disc; see, \textit{e.g.} Refs.~\cite{Hutsemekers:2010fw,Tinbergen:2003cp}.}

	We will not repeat here the discussion leading to our subsample ($p_{\mathrm{lin}}$ and $p_{\mathrm{circ}}$) or the method used to obtain our limits, as all the necessary details can be found in Ref.~\cite{Payez:2012vf}.
	Let us simply emphasise again that we have aimed at being as conservative as possible: taking only into account the influence of the magnetic field in the supercluster, allowing the magnetic field to get a longitudinal component most of the time (and allowing other quantities to fluctuate), as well as considering initially unpolarised light to make sure that the final polarisation due to the mixing is not overestimated for example.

	Using a conservative method leading to robust results, we can safely say that the mixing of light with axion-like particles is strongly constrained by the good-quality measurements of polarisation, and especially of circular polarisation, even when bandwidth effects are taken into account~\cite{Payez:2011sh}. However this might say something about the magnetic fields or the electron density, and not about ALPs.
	This is why we provide here the evolution of our limit with the maximum transverse magnetic field strength and with the mean electron density.

	The only reported detection available in the literature for the magnetic field in the local supercluster plane favours a collection of $\sim2~\mu$G magnetic field domains that are coherent over $\sim100$~kpc~\cite{Vallee:2011zz}, averaging to a field 5--10 times weaker at the supercluster scale.
	If we use this result, the 2$\sigma$-limit we obtain is $g<2.5\times10^{-13}$~GeV$^{-1}$ for $m<4.2\times10^{-14}$~eV; as emphasised in our paper~\cite{Payez:2012vf}, this limit can however be easily rescaled for any other value of the magnetic field strength as it always appears together with the coupling constant in the equations.

\begin{figure}[ht]
		\centerline{\includegraphics[width=\textwidth]{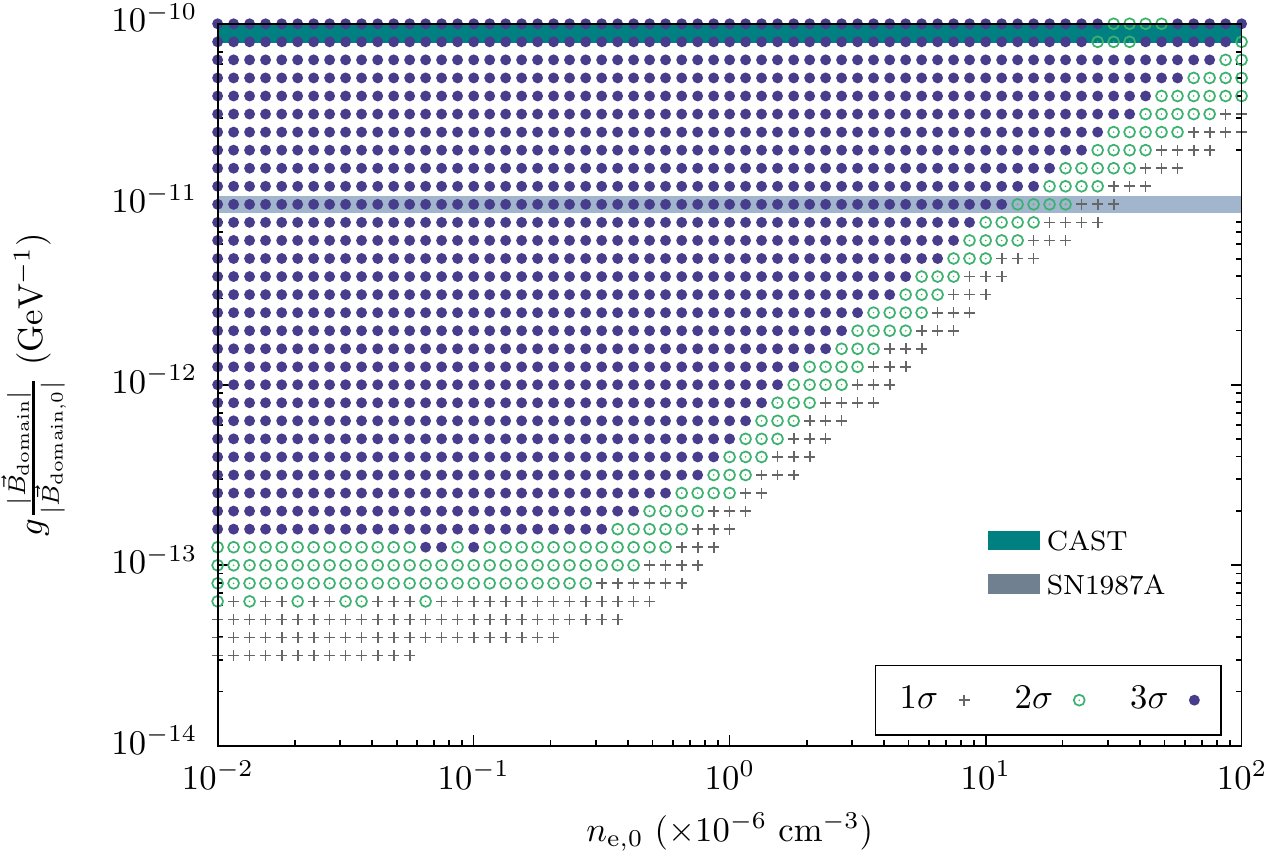}}
		\caption{Dependence of the plateau corresponding to $m^2\lll{\omega_{\mathrm{p}}}^2$ on the average electron density. The maximum transverse magnetic strength used here is $|\vec{B}_{\mathrm{domain},0}|=2~\mu$G.}\label{fig:AP_electrondensity}
		\label{sec:figures}
	\end{figure}

	On the other hand, in order to give the evolution of our limit with the average electron density, we can take advantage of the fact that, for ALP masses much smaller than the plasma frequency
	\begin{equation}
		\omega_{\mathrm{p}}=3.7\times10^{-14}~\textrm{eV}\times\sqrt{\frac{n_{\mathrm{e}}}{10^{-6}~\textrm{cm}^{-3}}},
	\end{equation}
	the two dimensionless quantities that determine the evolution of the Stokes parameters then become independent of $m$, leading to a plateau~\cite{Payez:2012vf}.
	Here, we give in Fig.~\ref{fig:AP_electrondensity} the evolution of this plateau for various values of the average electron density $n_{\mathrm{e},0}$; the case illustrated in the paper~\cite{Payez:2012vf} is given by $n_{\mathrm{e},0}=10^{-6}$~cm$^{-3}$, which is the value usually considered in superclusters. The information is summarised by saying that points in the parameter space are excluded at 1$\sigma$, 2$\sigma$ and 3$\sigma$ when the average probability that they contradict the observations is respectively 68.3\%, 95.5\% and 99.7\%.
	As anticipated in the paper, the limits on the coupling constant $g$ for nearly massless pseudoscalars would be more stringent for values of the plasma frequency smaller than the one we have considered, which reflects the fact that the mixing then becomes more efficient. Conversely, if the electron density in the local supercluster were to actually be much higher than that, then no constraint could be put on such particles as the mixing would then simply not take place in that medium.

\section*{Acknowledgments}

	We thank the IFPA group for the financial support received to attend this workshop, and acknowledge useful discussions with Jean-Ren\'e Cudell and Damien Hutsem\'ekers.


\begin{footnotesize}

\end{footnotesize}



\begin{thebibliography}{99}

\bibitem{Bahre:2013ywa}
  R.~B\"ahre, B.~D\"obrich, J.~Dreyling-Eschweiler, S.~Ghazaryan, R.~Hodajerdi, D.~Horns, F.~Januschek and E.-A.~Knabbe {\it et al.},
  JINST {\bf 1308} (2013) T09001
  [arXiv:1302.5647 [physics.ins-det]].

\bibitem{Vogel:2013bta}
  J.~K.~Vogel, F.~T.~Avignone, G.~Cantatore, J.~M.~Carmona, S.~Caspi, S.~A.~Cetin, F.~E.~Christensen and A.~Dael {\it et al.},
  arXiv:1302.3273 [physics.ins-det].

\bibitem{Payez:2013lwa}
  A.~Payez,
  Ph.D. thesis, arXiv:1308.6608 [astro-ph.CO].
  See the Introduction and Chap.~1.

\bibitem{Baker:2013zta}
  K.~Baker, G.~Cantatore, S.~A.~Cetin, M.~Davenport, K.~Desch, B.~D\"obrich, H.~Gies and I.~G.~Irastorza {\it et al.},
  Annalen Phys.\  {\bf 525} (2013) A93
  [arXiv:1306.2841 [hep-ph]].

\bibitem{Ringwald:2012hr}
  A.~Ringwald,
  Phys.\ Dark Univ.\  {\bf 1} (2012) 116
  [arXiv:1210.5081 [hep-ph]].

\bibitem{Payez:2012vf} 
  A.~Payez, J.R.~Cudell and D.~Hutsem\'ekers,
  JCAP {\bf 1207} (2012) 041
  [arXiv:1204.6187 [astro-ph.CO]].

\bibitem{Maiani:1986md}
  L.~Maiani, R.~Petronzio and E.~Zavattini,
  Phys.\ Lett.\ B {\bf 175} (1986) 359.

\bibitem{Raffelt:1987im}
  G.~Raffelt and L.~Stodolsky,
  Phys.\ Rev.\ D {\bf 37} (1988) 1237.

\bibitem{Harari:1992ea}
  D.~Harari and P.~Sikivie,
  Phys.\ Lett.\ B {\bf 289} (1992) 67.

\bibitem{Sikivie:1983ip}
  P.~Sikivie,
  Phys.\ Rev.\ Lett.\  {\bf 51} (1983) 1415
   [Erratum-ibid.\  {\bf 52} (1984) 695].

\bibitem{HornsAlighieri}
  D.~Horns, L.~Maccione, A.~Mirizzi and M.~Roncadelli,
  Phys.\ Rev.\ D {\bf 85} (2012) 085021
  [arXiv:1203.2184 [astro-ph.HE]];
  S.~d.~S.~Alighieri, F.~Finelli and M.~Galaverni,  
  Astrophys.\ J.\  {\bf 715} (2010) 33
  [arXiv:1003.4823 [astro-ph.CO]].
  
\bibitem{Hutsemekers:2010fw}
  D.~Hutsem\'ekers, B.~Borguet, D.~Sluse, R.~Cabanac and H.~Lamy,
  Astron. Astrophys. {\bf 520} (2010) L7
  [arXiv:1009.4049 [astro-ph.CO]].

\bibitem{Courtois:2013yfa}
  H.~M.~Courtois, D.~Pomar\`ede, R.~B.~Tully, Y.~Hoffman, and D.~Courtois,
  Astron. J. {\bf 146} (2013) 69
  [arXiv:1306.0091 [astro-ph.CO]].

\bibitem{Tinbergen:2003cp}
  J.~Tinbergen,
  Astrophys. Space Sci.  {\bf 288} (2003) 3.


\bibitem{Payez:2011sh} 
  A.~Payez, J.R.~Cudell and D.~Hutsem\'ekers,
  Phys.\ Rev.\ D {\bf 84} (2011) 085029
  [arXiv:1107.2013 [astro-ph.CO]].

\bibitem{Vallee:2011zz}
  J.~P.~Vall\'ee,
  New Astron.\ Rev.\  {\bf 55} (2011) 91.

\end{thebibliography}
\end{document}